# Abundant Circumstellar Silica Dust and SiO Gas Created by a Giant Hypervelocity Collision in the ~12 Myr HD172555 System

C.M. Lisse[1], C. H. Chen[2], M. C. Wyatt[3], A. Morlok[4],
I. Song[5], G. Bryden[6], and P. Sheehan[7]

[1] JHU-APL, 11100 Johns Hopkins Road, Laurel, MD 20723 carey.lisse@jhuapl.edu.

[2] STScI 3700 San Martin Dr. Baltimore, MD 21218 cchen@stsci.edu.

[3] Institute of Astronomy, University of Cambridge, Madingley Road, Cambridge CB3 0HA, UK wyatt@ast.cam.ac.uk.

[4] Dept. of Earth and Planetary Sciences, Faculty of Science, Kobe University, Kobe 657-8501, Japan
Current address : CRPG-CNRS, UPR2300, 15, rue Notre Dame des Pauvres, BP20, 54501 Vandoeuvre les Nancy, France amorlok@crpg.cnrs-nancy.fr

[5] Department of Physics and Astronomy, The University of Georgia, Athens, GA 30602, song@uga.edu

[6] Jet Propulsion Laboratory, 4800 Oak Grove Drive, Pasadena, CA 91109 Geoffrey.Bryden@jpl.nasa.gov

[7] Department of Physics and Astronomy, University of Rochester, Rochester, NY
psheeha2@mail.rochester.edu

Proposed Running Title: **Silica and SiO From a Hypervelocity Collision in HD172555**

Please address all future correspondence, reviews, proofs, etc. to:

Dr. Carey M. Lisse

Planetary Exploration Group, Space Department

Johns Hopkins University, Applied Physics Laboratory

11100 Johns Hopkins Rd

Laurel, MD 20723

240-228-0535 (office) / 240-228-8939 (fax)

Carey.Lisse@jhuapl.edu

# ABSTRACT

The fine dust detected by IR emission around the nearby β Pic analogue star HD172555 is very peculiar. The dust mineralogy is composed primarily of highly refractory, non-equilibrium materials, with approximately three-quarters of the Si atoms in silica ($SiO_2$) species. Tektite and obsidian lab thermal emission spectra (non-equilibrium glassy silicas found in impact and magmatic systems) are required to fit the data. The best-fit model size distribution for the observed fine dust is $dn/da = a^{-3.95 \pm 0.10}$. While IR photometry of the system has stayed stable since the 1983 IRAS mission, this steep a size distribution, with abundant micron-sized particles, argues for a fresh source of material within the last 0.1 Myr. The location of the dust with respect to the star is at $5.8 \pm 0.6$ AU (equivalent to $1.9 \pm 0.2$ AU from the Sun), within the terrestrial planet formation region but at the outer edge of any possible terrestrial habitability zone. The mass of fine dust is $4 \times 10^{19} - 2 \times 10^{20}$ kg, equivalent to a 150 – 200 km radius asteroid. Significant emission features centered at 4 and 8 μm due to fluorescing SiO gas are also found. Roughly $10^{22}$ kg of SiO gas, formed by vaporizing silicate rock, is also present in the system, and a separate population of very large, cool grains, massing $10^{21} - 10^{22}$ kg and equivalent to the largest sized asteroid currently found in the Solar System's main asteroid belt, dominates the solid circumstellar material by mass. The makeup of the observed dust and gas, and the noted lack of a dense circumstellar gas disk, strong primary x-ray activity, or an extended disk of β meteroids argues that the source of the observed circumstellar materials is a giant hypervelocity (> 10 km sec$^{-1}$) impact between large rocky planetesimals, similar to the ones which formed the Moon and which stripped the surface crustal material off of Mercury's surface.

1. **Introduction**

**HD172555 (HR7012)** is an A5V star (V = 4.8, Michigan Spectral catalogue I, Houk and Cowley 1975) at 29.2 pc from the Earth, and as a member of the β Pic moving group, is estimated to be very young, ~12Myr (Zuckerman and Song 2004). The star system HD172555 was first recognized by Schutz *et al.* (2005) in their ESO study of Southern IRAS IR bright stars as having a very high ratio of mid-IR to total emission, $f_{IR}/f_{Bol}$ = $5 \times 10^{-4}$, large even when compared to other A stars with known circumstellar disks. It was also recognized by Chen *et al.* (2006) as being one of four systems in their Spitzer survey of IRAS-discovered debris disks with pronounced mid-IR spectral features (Fig 1a).

The parallels to the A5 β Pic system (at 19 pc) distant are also quite striking: HD172555 is co-moving (and thus probably coeval) with β Pic, is a similar class A5 star, and their B-V colors and $M_v$ are the same within 0.05 magnitudes (Moór *et al.* 2006; Figure 1b). β Pic is well known for its highly structured circumstellar disk of dusty material, extending out to 100's of AU from the primary. Current work suggests the β Pic material has at least 3 planets in the throes of formation (Okamoto *et al.* 2004), clumps of material undergoing collisions (Telesco *et al.* 2005), and a composition that is very rich in carbon (Roberge *et al.* 2006). By implication, HD172555 should also have an extensive circumstellar disk. In an HST/ACS coronagraphic mode search for scattered light from circumstellar dust around members of the β Pic moving group (Ardila *et al.* 2009, in preparation), no extended emission was detected around HD172555 to the limits of the size of the coronagraphic disk (0.5" radius or 15 AU at 29 pc) – so any significant circumstellar material must be localized inside 15 AU. The HD172555 system also contains a wide binary companion, an M0 star, CD -64 1208 at ~2000 AU (1.1' at 29 pc; Feigelson et al. 2006). While CD -64 1208 may dynamically limit the extent of any very large dust orbits with semimajor axes greater than 1000 AU, it is unimportant for the warm dust under study here, nor would it effect the dynamics of planetary objects found in orbits analogous to the known major bodies of our solar system.

The overall spectral energy distribution (Figures 1 and 3) for the circumstellar material orbiting HD172555 has a best-fit blackbody temperature of ~245K, which means the dust is relatively warm and located at around 3 to 4 AU from the primary, consistent with the HST imaging results. (Note that the distant companion's insolation is not important with respect to the energy budget of the warm dust.) 'Warm' circumstellar dust with $T > 150$ K is found in only 5-10% of A-star systems (Su *et al.* 2006, Morales *et al.* 2009), and is a hallmark for abundant particles within a few AU of the central star, implying that something special is going on in the system. Unless the star is extremely young, less than a few Myr old, the warm dust is also not primordial, and is proof that a reservoir of dusty material (asteroids, comets, planets, moons, Kuiper Belt Objects, etc.) is present. Given HD172555's estimated age of 10 -12 Myr, close to the estimated age for the initiation of terrestrial planet formation (Wetherill 1990; Yin *et al.* 2002; Chambers 2004), and its similarity to the fascinating β Pic system, the system is worthy of study for evidence of early solar system formation processes.

In this paper we analyze in detail the Spitzer IRS spectral measurements, utilizing strong features in the mid-IR spectrum to allow us to determine the major dust components producing the observed emission. We used a similar methodology to study the amount, kind, and location of the dust and water gas/ice excavated from the comet 9P/Tempel 1 (Lisse *et al.* 2006), the dust and water ice found in comet C/1995 O1 (Hale-Bopp) and the comet dominated primordial disk around the Herbig A0 star HD100546 (Lisse *et al.* 2007a), in the dense zody cloud around the 2-10 Gyr old K0V star HD69830 (Lisse *et al.* 2007b), in near orbit around the ancient DZ white dwarf GD29-38 (Reach *et al.* 2008), and in the terrestrial planet forming region around the young F5 star HD113766 (Lisse *et al.* 2008). While we were able to obtain a good match to the HD172555 observed spectrum, we find that unlike for any of the previously studied systems, there is a very unusual $SiO_2$ (silica) dust and SiO gas composition in the HD172555 circumstellar material.

## 2. Observations & Methodology

**2.1 Observations.** For their IRS observations of HD172555, Chen *et al.* (2006) utilized a combination of the Short-Low (5.2–14.0 µm, $\lambda/\Delta\lambda$ ~ 90), Short-High (9.9–19.6 µm, $\lambda/\Delta\lambda$ ~ 600), and Long-High (18.7-37.2 µm, $\lambda/\Delta\lambda$ ~ 600) modules. The reduction and analysis of the spectra was conducted with the *Spitzer* IRS instrument team's SMART program, V15 (Higdon *et al.* 2004). In order to avoid time-consuming peak-up, the observatory was operated in IRS spectral mapping mode where a 2 × 3 raster (spatial × dispersion) centered on the star was performed (Watson *et al.* 2004). In order to verify the presence of the 8 µm SiO feature in the data (§ 2 and 3), the spectra were re-extracted twice, with each nod analyzed to ensure the sharp feature was not a detector effect. A total of 1384 independent spectral points were obtained over the range 5.2-36 µm. Relative calibration of the spectral orders, and fringing in the long wavelength data were important issues that had to be dealt with in treating the data. We refer the reader to Chen *et al.* (2006) for more details of the IRS raw data reduction.

The 5-35 µm HD172555 disk excess flux studied in this work was then calculated by removing the stellar photospheric contribution from the IRS spectrum of Chen *et al.* (2006). The photospheric contribution was modeled by assuming that the HD172555 spectrum is represented by an A5 member of the Beta Pic moving group with an age of 12 Myr (Zuckerman and Song 2004). The star was assumed to have solar abundance, log $g = 4.5$, and $E(B − V) = 0.01$ (determined using the Cardelli *et al.* 1989 extinction law). The stellar photospheric flux was then estimated by minimum $\chi^2$ fitting of the appropriate Kurucz stellar atmosphere model to optical and 2MASS (0.3 – 3 um) archival photometry[1]. The photosphere-removed flux is presented and compared to other significant mid-IR dust spectra (Figure 2), to the emission from glassy silica powder (Figure 3), and to the Deep Impact compositional model in Figure 4.

## 2.2    The Deep Impact Tempel 1 Dust Model

To understand the information derived from the HD172555 excess IRS spectra, we summarize here the relevant portions of the Deep Impact experiment, its Spitzer IRS measurement, and the Tempel 1 Dust Model created to interpret the measurements.

---

[1] See URL http:// http://nsted.ipac.caltech.edu/

Further details of the spectral analysis are described in the literature in the Supplementary Online material for Lisse *et al.* 2006, and in the main text of Lisse *et al.* 2007a.

**2.2.1 Deep Impact.** The recent Deep Impact hypervelocity experiment, one of the few direct man-made astrophysical experiments on record, produced a mix of materials - volatilized cometary material, silicaceous droplets, and excavated, largely unprocessed bulk material – after a small bolide of ~370 kg mass impacted the nucleus of comet 9P/Tempel 1 (hereafter Tempel 1) at 10.2 km sec$^{-1}$ on 2005 July 4, when the comet was 1.51 AU from the Sun (A'Hearn *et al.* 2005, Melosh 2006, Richardson *et al.* 2007). Due to the very low bulk modulus (< 10 kPa) and escape velocity (~ 1 m sec$^{-1}$) of the nucleus, the ejected material was heavily dominated, by 2-3 orders of magnitude in mass, by the excavation of bulk unprocessed material, as were the Spitzer emission spectra. This bulk material was excavated from the nucleus from depths as large as 30m and was largely unaltered, except for disruption of loosely-held macroscopic fractal particles into the individual sub-fractal, micron-sized components, as demonstrated by the presence of significant amounts of non-refractory water ice in the ejecta (Lisse *et al.* 2006; Sunshine *et al.* 2006). *Spitzer* IRS 5–35 μm spectra were taken within minutes of the impact, both before and after. The ejected material cooled from effects due to the impact within seconds to minutes, and separation of the ejecta emission spectrum from blackbody emission at LTE due to large, optically thick dust particles in the ambient coma was easily made. The resulting highly structured spectrum of the ejecta showed over 16 distinct spectral features at flux levels of a few Janskys (Lisse *et al.* 2006) that persisted for more than 20 hours after the impact.

**2.2.2 Thermal Emission.** The emission flux from a collection of dust is given by

$$F_{\lambda,\text{mod}} = \frac{1}{\Delta^2} \sum_i \int_0^\infty B_\lambda(T_i(a,r_*)) Q_{abs,i}(a,\lambda) \pi a^2 \frac{dn_i(r_*)}{da} da$$

where $T$ is the particle temperature for a particle of radius $a$ and composition $i$ at distance $r_*$ from the central star, $\Delta$ is the distance from *Spitzer* to the dust, $B_\lambda$ is the blackbody radiance at wavelength $\lambda$, $Q_{abs}$ is the emission efficiency of the particle of composition $i$ at wavelength $\lambda$, $dn/da$ is the differential particle size distribution (PSD) of the emitted dust, and the sum is over all species of material and all sizes of particles for the dust. Our spectral analysis consists of calculating the emission flux for a model collection of dust, and comparing the calculated flux to the observed flux. The emitted flux depends on the composition (location of spectral features), particle size (feature to continuum contrast), and the particle temperature (relative strength of short versus long wavelength features), and we discuss each of these effects below.

**2.2.3 Composition.** To determine the mineral composition the observed IR emission is compared with the linear sum of laboratory thermal infrared emission spectra. As-measured emission spectra of randomly oriented, μm-sized powders were utilized to directly determine $Q_{abs}$. The material spectra were selected by their reported presence in interplanetary dust particles, meteorites, in situ comet measurements, YSOs, and debris disks (Lisse *et al.* 2006). By building up a spectral library, we tested for the presence of over 80 different species in the T1 ejecta, in order to approach the problem in an unbiased fashion. We also expected, given the large number of important atomic species in astrophysical dust systems (H, C, O, Si, Mg, Fe, S, Ca, Al, Na, …) that the number of important species in the dust would be on the order of ~10.

The list of materials tested included multiple silicates in the olivine and pyroxene class (forsterite, fayalite, clino- and ortho-enstatite, augite, anorthite, bronzite, diopside, and ferrosilite); phyllosilicates (such as saponite, serpentine, smectite, montmorillonite, and chlorite); sulfates (such as gypsum, ferrosulfate, and magnesium sulfate); oxides (including various aluminas, spinels, hibonite, magnetite, and hematite); Mg/Fe sulfides (including pyrrohtite, troilite, pyrite, and niningerite); carbonate minerals (including calcite, aragonite, dolomite, magnesite, and siderite); water-ice, clean and with carbon dioxide, carbon monoxide, methane, and ammonia clathrates; carbon dioxide ice; graphitic and amorphous carbon; and the neutral and ionized PAH emission models of Draine & Li (2007).

A model phase space search easily ruled out the presence of a vast majority of our library mineral species from the T1 ejecta. Only convincing evidence for the following as the majority species in the Tempel 1 ejecta was found (Lisse et al. 2006): crystalline silicates like forsterite, fayalite, ortho-enstatite, diopside, ferrosilite and amorphous silicates with olivine and pyroxene like composition; phyllosilicates similar to nonerite; sulfides like ningerite and pyrrohtite; carbonates like magnesite and siderite; water gas and ice; amorphous carbon (and potentially native Fe:Ni; and ionized PAHs. This list of materials compares well by direct comparison to numerous in situ and sample return measurements (e.g., the Halley flybys and the STARDUST sample return; see Lisse *et al.* 2007a for a detailed list). Before HD172555, these 7 classes of minerals (Ca/Fe/Mg-rich silicates, carbonates, phyllosilicates, water-ice, amorphous carbon, ionized PAHs, and Fe/Mg sulfides; 15 species in all) were successfully used to model thermal emission from dust emitted by 6 solar system comets, 3 extra-solar YSOs, and two mature exo-debris disks (Lisse et al. 2006, 2007a, 2007b, 2008).

**2.2.4 Particle Size Effects.** Particles of 0.1–1000 um are used in fitting the 5–35 um data (although results to date have shown a sensitivity only to the 0.1–20 µm particle size range), with particle size effects on the emissivity assumed to vary as

$$1 - Emissivity(a, \lambda) = [1 - Emissivity(1um, \lambda)]^{(a/1um)}$$

The particle size distribution (PSD) is fit at log steps in radius, i.e., at [0.1, 0.2, 0.5, 1, 2, 5,…100, 200, 500] µm. Particles of the smallest sizes have emission spectra with very sharp features, and little continuum emission; particles of the largest sizes are optically thick, and emit only continuum emission. Both the T1 ejecta spectrum and the HD172555 spectrum show strong, sharp, high contrast features, indicative of the presence of small (~1 µm) grains. For the Tempel 1 ejecta, the PSD was found to be unusually narrow, consisting predominantly of 0.5–2.0 µm particles (Lisse *et al.* 2006). For the HD172555 excess, a steep power law spectrum of particle sizes, dominated by small grains, was found to be necessary to fit the Spitzer data. New for this work was the additional requirement of including a separate population of large cool grains with a > 100 μm.

**2.2.5 Particle Temperature.** Dust particle temperature is determined at the same log steps in radius used to determine the PSD. Particle temperature is function of particle composition and size at a given astrocentric distance. The highest temperature for the smallest particle of each species is free to vary, and is determined by the best-fit to the data; the largest, optically thick particles (1000 µm) are set to the LTE temperature, and the temperature of particles of in between sizes is interpolated between these extremes by radiative energy balance. We model the unresolved dust excess around HD172555 as a relatively localized dust torus, and a disk as a linear sum of individual torii.

The T1 ejecta were, by experimental design, all highly localized at 1.51 AU from the Sun. We use this fact to *empirically* determine the effective distance of the emitting material from HD172555. To do this, the best-fit model temperature for the smallest and hottest (0.1 -1 µm) particles for each material from our analysis is compared to the temperatures found for the (0.5 - 2.0 µm) particles of the Tempel 1 ejecta (Lisse *et al.* 2006), using the relation

$$T_{dust} = T_{T1ejecta}(L_*/L_{solar})^{1/4}(1.51 AU/r_*)^{1/2}$$

where $T_{dust}$ is the temperature of the dust around HD172555, $T_{T1\ ejecta}$ are as given in Lisse *et al.* 2006 (~ 340 K), $L_*$ = bolometric luminosity of HD172555, and r* is the distance of the dust particle from HD172555.

**2.2.6 Model Summary.** Our method has limited input assumptions, uses physically plausible emission measures from randomly oriented powders, rather than theoretically derived Mie values, and simultaneously minimizes the number of adjustable parameters. The free parameters of the model are the relative abundance of each detected mineral species, the temperature of the smallest particle of each mineral species, and the value of the particle size distribution at each particle size (Table 1). Best-fits are found by a direct search through (composition, temperature, size distribution) phase space. The total number of free parameters in the model used to fit the IRS HD172555 spectrum = 11 relative abundances + 11 hottest particle temperatures + 1 power law size index = 23. (Since the results of our modeling show we could describe the dust temperature using 2 parameters, in application we actually use only 14 adjustable values.)

The model has allowed us to get beyond the classical, well known olivine-pyroxene-amorphous carbon composition to the second-order, less emissive species like water, sulfides, PAHs, phyllosilicates, and carbonates. We are able to determine the overall amounts of the different major classes of dust-forming materials (olivines, pyroxenes, sulfides, water, etc.) and the bulk elemental abundances for the most abundant atoms in these materials (H, C, O, Si, Mg, Fe, S, Ca, Al). Applying our analysis, with a series of strong 'ground truth' checks of its validity, is highly diagnostic for interpreting mid-IR spectra of distant dusty systems like YSOs, debris disks, and PNs.

**2.3 Required Deviations From the 'Deep Impact Standard Compositional Model'.**
The main difference between the approach used to model the mid-infrared spectrum of HD172555 and that used to decompose previously studied dusty cometary and exo-systems is that the usual linear sum of laboratory thermal infrared emission spectra do not fit the HD172555 spectrum. I.e., no combination of emission from the Fe/Mg olivines, Ca/Fe/Mg pyroxenes, Fe/Mg sulfides, phyllosilicates, amorphous inorganic carbon, water ice/gas, PAHs, and carbonates commonly found in circumstellar dust, or from the other 80 species in our spectral library was able to reproduce the observed spectrum. A host of new mineral species were studied and compared to the HD172555 spectrum. The majority of these species focused on matching the pronounced emission feature from 7.5 - 10 um, in the usual range for an Si-O vibration mode. Lunar species such as anorthite and basalt, and all the allotropes of silica available in the literature were studied : quartz, cristoblite, tridymite, amorphous silica, protosilicate, obsidian, and various tektite compositions. Also examined were mineral sulfates (such as gypsum, ferrosulfate, and magnesium sulfate) and oxides (including various aluminas, spinels, hibonite, magnetite, and hematite).

We found the high SNR of the Spitzer data in the 7 - 13 $\mu$m region to be highly constraining on the possible species present. A clear match to the Spitzer spectrum at the 95% confidence level (C.L.) was found only in the case of a spectrum dominated by emission from the glassy silicas: obsidian and tektites. The presence of obsidian and tektite-like material, produced on Earth as the kinetic product of very high temperature processing of silicaceous materials in lavas and impact craters (i.e., by quick quenching of rapidly melted rocks) was immediately suggestive of a violent origin for the circumstellar material found around HD172555.

Just as interesting was the discovery of 2 sharp residuals in the emission after all the solid state emission features had been removed. These features were found with good fidelity at 7.5 - 10 µm and in a much noisier feature rising towards short wavelengths at 5-6 µm (Figure 4, right panel). We could not find any solid state material in the literature that matched these residuals. On the other hand, a literature search of possible gas species produced an excellent match with the fundamental ro-vibrational linear stretch complex of the SiO molecule – it has P, Q, and R branches at 7.5 - 10 um, centered at ~8 um. The first overtone of the SiO stretch lies at ~ 4 um, and the ro-vibrational manifolds have transitions extending from 3 – 6 um. Examination of the transition moment calculations of Drira *et al.* (1997) and the IRAS/LRS and ISO-SWS measurements of Alpha Tau, a oxygen-rich K-star on the asymptotic giant branch with a known SiO absorption feature (Cohen and Davies 1995) demonstrated a good fit to the residuals. Once we adopted an opacity for SiO gas based on ISO-SWS observations of Alpha Tau, we found that we could easily account for the sharp fall in emissivity at 5 - 6.5 um, the extremely flat emissivity at 6.5-7.5 um, and the sharp rise just longwards of 7.5 µm. SiO gas is the species created when olivines and pyroxenes are vaporized in high temperature processes - the minerals decompose into SiO and metal (FeO, MgO, CaO, AlO, etc.) oxides. The same processes can form silica from the olivine and pyroxene silicates.

A third component we have not had to use in our previous analyses, but find necessary to obtain a good fit to the Spitzer data (especially at the longer wavelengths) is material contained in a very large particle dust population ( a > 100 µm), radiating as blackbodies at ~200K. As described in Sec 2.2.4, for all previous analyses of Spitzer bright dusty disk spectra we have been able to fit both the sharp features and the continuum of the spectrum with a power law size distribution of dust, 0.1 < a < 1000 um. At 80% relative surface area of all the other solid dust components, the large dust particle population is clearly important in the overall mix of materials. It is, however, difficult to characterize due to its relative lack of spectral structure. No mineralogical information can be derived, and only a lower limit for the mass.

## 3. Results

**3.1 Dust Composition.** As stated in §2, what we have found for the circumstellar dust around HD172555 is a mineralogy unlike anything else we have examined (Figure 2). The Si-O vibrational stretch region of the spectrum (~8-13 um), normally dominated by the emission from ferromagnesian silicates at 9.5 - 11.5 um, has a pronounced feature at 7.5 to 10 µm instead (Figure 3). There is a subsidiary peak due to olivine at 11.2 um, and some smaller features at ~20 and 33 µm also probably due to olivine, usually found in astrophysical spectra of YSOs and comets (Figure 2). There is also some evidence for pyroxene in the dust spectra, with emission peaking at 9.5 - 11 um. What is totally new, though, is the very strong peak at 9.3 um, and the sharp shoulder at 7.5 - 8.5 µm (most clearly seen as a 'step' in Figs 2-4). Olivines and pyroxenes do not have substantial emission features in this range. We can rule out massive amounts of PAHS as the source of the emission, by the lack of strong isolated features at 6.2, 7.8 um, and 8.6 um, by comparison to the models of Li and Draine (2008), and by direct comparison to the PAH-rich HD100546 system (Figure 2).

The only major planetary mineralogical component that we can find in the literature that is associated with a 9.3 μm feature is amorphous silica (not silicate; Figure 3), the kinetically favored species produced by intense, rapid heating and melting/vaporization of silicate rich bodies followed by quick quenching. E.g., silica rich materials are found in terrestrial lave flows and in the tektites and glasses found at the hypervelocity (v > 10 km sec$^{-1}$) impact craters on the Earth and Moon (Warren 2008) and in some chondritic meteorite material (Tissandier *et al.* 2002), as large sphericules of refractory material condensed from the proto-planetary disk. In our model fits, a good match to HD172555 9.3 μm emission feature is found **only** with powdered obsidian and powdered tektite material, i.e. glassy silica. These materials also contain minority fractions of MgO, FeO, CaO, $Al_2O_3$, $Na_2O$, and $K_2O$ (Koeberl 1988).

Crystalline silica phases like cristobalite and tridymite do not fit the data well, and the 12.6 um feature found for T Tauri systems rich in crystalline silica (Sargent et al. 2006, Forrester private communication 2008) is not present. The matching of the IR emission features to amorphous silica is in contrast to the preliminary models of Chen *et al. 2006* for HD172555, who found cristobalite to be a good fit for the excess spectrum. While a 9 um silica feature is also present in the cristobalite spectrum, the location of the peak is too far to the short wavelength side to give a good fit to the HD172555 mid-infrared spectrum.

Our best-fit spectral decomposition for the HD172555 excess spectrum, with $\chi^2_\nu = 1.04$ (95% C.L. = 1.07 for 1186 d.o.f.) is presented in Figure 4 & Table 2, and the derived elemental abundances in Figure 5 & Table 3. From these, we note that the dust is very amorphous silica rich, and shows a strong residual at 7.5 - 8.2 µm, evidence for the presence of SiO gas. There is a deficit of pyroxene versus olivine. The silicates that are present are almost all crystalline. Unlike other young systems (Lisse *et al.* 2008), no silicates of amorphous pyroxene composition are present, and only a small amount of silicates of amorphous olivine composition were detected. The amorphous carbon content is small. No low temperature organic carbonaceous materials are present in sufficient quantities in the circumstellar dust to be detected in the infrared, and no water or water ice.

There are parallels between the fine silica dust composition derived here, and the reported mineralogical compositions of some chondrules (Brearley and Jones 1998) and Glass with Embedded Sulfide particles (GEMS; Bradley *et al.* 1999, 2005). However, the HD172555 system age of 10-12 Myr is a factor of 3-4 older than that derived for chondritic matter in the solar system.

The major atomic refractory species, with the exception of Al, are consistently depleted versus solar. Despite detailed searches in our data, there are no obvious features of FeO or MgO mineral species (expected in addition to SiO gas if olivine or pyroxene pyrolitic decomposition has occurred). If our derived composition is accurate, then we have to explain the high Si and O content in the circumstellar material. One likely explanation is that the parent body(s) of the dust contributed only non-solar abundance SiO-rich starting material, as from a highly differentiated crust. Another possibility is that whatever high temperature process created the silica and SiO gas preferentially smelted, or removed, SiO from an initially solar abundance rocky material. It is important to note, however, that our results are based on the assumption that we are accurately sampling a representative cross section of the atoms in the circumstellar material. Some atomic species may have preferentially condensed into large aggregates that are not readily detectable in the Spitzer spectrum due to their size (i.e., the large, cool blackbody particles).

**3.2 Dust Mass**: The best-fit model size distribution for the HD172555 circumstellar dust producing the sharp silica feature, $dn/da = a^{-3.95 \pm 0.10}$ with $0.1 < a < 1000$ µm (Figure 6), is more small particle dominated than a purely collisional equilibrium distribution of $dn/da = a^{-3.50}$ (Dohnanyi 1969). This steep a size distribution, with abundant micron-sized particles, argues for a fresh source of material within the last 0.1 Myr, created in a non-equilibrium fashion. Integrating the best-fit model PSD for the fine dust (Figure 6), we find an estimated lower mass limit (optically thin dust, 1000 µm largest particle) for the circumstellar dust in the system of $4 \times 10^{19}$ kg. This implies that the parent body of the fine dust, if asteroidal in nature with mean density of 2.5 g cm$^{-3}$, would be at least 150 km in radius. The dust mass increases roughly logarithmically with the upper particle radius assumed, i.e. the mass estimate would be $1 \times 10^{20}$ kg for a 10 m largest particle, implying at least a 200 km radius, 2.5 g cm$^{-3}$ parent body as the source of the material.

Additional mass is contained in a separate large particle dust population, the material radiating as blackbodies at ~200K. Unfortunately we have little information concerning their mass (or composition) due to the lack of spectral information. We note that given that the relative surface area in this population is significant, the relative amount of mass is likely to be important as well. However, the estimated mass depends very much on the actual particle size, and can vary by orders of magnitude depending on the size model assumed. A lower bound estimate to this dust mass can be obtained if we assume all the 200K particles are ~ 100 µm in radius, the smallest size which will act radiatively as pure blackbodies across the Spitzer IRS passband, and note that the relative surface area in large particles is roughly 80% that in the fine population. In this case we find that there is approximately 50 times more mass in the large cool particles than in the small fine particles, or $50 * [6 \times 10^{19} - 2 \times 10^{20}]$ kg = $[3 \times 10^{21} - 1 \times 10^{22}]$ kg. The equivalent asteroidal body required to provide this amount of dust is about the size and mass of Ceres (Table 4). For the sake of argument, and supported by the analysis in the next paragraph, we adopt this range of masses for the large dust component for the rest of this work.

Another method of estimating the total mass of circumstellar dust in HD172555 is derived from the strong similarity between the spectral signatures of the HD172555 and the low mass FN Tau T-Tauri Star (TTS) disk in the mid-IR (Section 4.2). The 10 µm emission level of 0.3 Jy at ~140 pc for FN Tau, compared to the 10 µm emission of 0.9 Jy at 29 pc for HD172555, implies that for the same kind of dust (i.e., same emissivity/unit mass material), there is ~8x more mass in the FN Tau circumstellar disk than in the HD172555 system. Using Kudo *et al.*'s (2008) estimate of $2 \times 10^{-3}$ $M_\oplus$ of warm dust in the disk inside 30 AU, we derive for FN Tau a mass of $1 \times 10^{22}$ kg. Applying the factor of 8 reduction for the observed fluxes, this implies ~$2 \times 10^{21}$ kg of dust in HD17255, roughly consistent with the sum total of $10^{21}$ - $10^{22}$ kg estimated above for the mass of fine and cold large dust in the system.

We estimate the amount of SiO gas detected following Crovisier (2002). The g-factor for SiO can then be calculated using the value for the Einstein A coefficient of 6.6 ($\Delta v = 1$, 1-0 transition, Drira *et al.* 1997), assuming an SiO molecule fluorescing at 1 AU due to photons originating in an HD172555 A5 photosphere of 8000°K color temperature :

$$g = 5.43 \times 10^{-6}*(R_*/R_\odot)*6.6/(e^{[hc/\lambda kT]} -1)$$

$$= 5.43e\text{-}6*6.6*/(e^{[6.626 \times 10^{-34} * 3.0 \times 10^{14}/8.0/1.38 \times 10^{-23}/8000]}-1)$$

$$= 1.4 \times 10^{-4} (R_*/R_\odot)$$

Assuming $R_*/R_\odot$ for an A5 star is 1.6 (Lacy 1977), the number of molecules of SiO gas can be calculated in a straightforward manor from the value of the g-factor, the observed strength of the IR line, assuming the stellar flux falls off as $1/r_h^2$:

$$N_{mol\ SiO} = 4\pi \Delta^2 r_h^2 * F_\nu\ d\lambda / (g * h\lambda)$$

$$= 4\pi*(29\ pc*3.1\times10^{16} m\ pc^{-1})^2*(5.8\ AU)^2*(0.07 \pm 0.02\ (2\sigma)\ Jy * 10^{-26}\ Wm^{-1}Hz^{-1}Jy^{-1})$$

$$* 0.7 \pm 0.2\mu m\ (2\sigma)\ /(1.4\times10^{-4} * 1.6 * 6.63\times10^{-34}\ J\ sec * 8\mu m)$$

$$= 1.4 \pm 0.7\ (2\sigma)\ \times10^{47}\ molecules$$

where F is the flux in the line, $d\lambda$ is the width of the line, $\lambda$ is the center wavelength of the band, g is the g-factor for the band, $N_{mol}$ is the number of molecules in the beam, $\Delta$ is the Spitzer - HD172555 distance, and $r_h$ is the SiO molecule-HD172555 distance. Converting from the number of SiO gas molecules to the mass of SiO gas, assuming 44 amu per SiO gas molecule, we find

$$M_{SiO} = N_{mol\ SiO} * 44\ amu * 1.66 \times 10^{-27}\ kg\ amu^{-1}$$

$$= 1.1 \pm 0.6\ (2\sigma) \times 10^{22}\ kg\ SiO\ gas$$

Comparing this gas mass estimate to the $10^{19}$ - $10^{20}$ kg of fine dust, the $\geq 3 \times 10^{21}$ kg of coarse dust, and the 1-2 x $10^{22}$ kg of SiO gas, we see that the total mass of material is on the order of the mass of the Moon or Mercury, and that most of the mass of the circumstellar silicaceous material is in the gas phase and the large cold dust.

There are a number of energetic early solar system physical processes (Section 4.2), like nebular gas shocks and stellar flares, that could lead to such a mix of material, by preferentially vaporizing the small end of an initial dust population, leaving behind a remnant large dust population and a slowly recondensing, new fine dust population. An energetic giant impact could also form such a mix, with the SiO gas formed by direct vaporization of rocky material at the surface of contact, tektite and obsidian-like material from solidification of molten material spalled from the impact site, and large pieces of solid material excavated and ejected relatively intact from the edges of the impact site by induced shock waves.

**3.3 Dust Temperature and Location:** The best-fit blackbody function to the total 5 - 33 $\mu$m HD172555 dust excess spectrum has $T_{bb}$ = 335 K. While extremely useful for quickly removing the zeroth-order effects of dust temperature on the as-observed spectrum, a blackbody fit is a gross simplification, lumping in the emission behavior of multiple chemical species and particle sizes. Our more detailed compositional modeling finds a maximal temperature for the smallest silicaceous dust grains of 305 K (compare to 260K estimate from IRAS photometry, and a 520K + 170K estimate from Chen *et al.* 2006) for a primary with $L_*$ = 9.5 $L_{Sun}$(Wyatt *et al.* 2007b, Table 1). To determine where the dust lies with respect to the primary, we scale the results of the Deep Impact experiment, where fine micron sized dust was released at 1.51 AU into the Sun's radiation field and promptly observed by the Spitzer IRS (Sec 2.2.5). This technique has worked well in comparison to imaging measures of the dust location. E.g., we placed in the location of the dominant emitting dust from the inner cavity wall of the HD100546 disk at 13 AU (Lisse *et al.* 2007a), and STIS 1-D spectroscopic measurements find a cavity for this disk of radius 13.2 AU across (Grady *et al.* 2001, 2005). Also, our best-fit model for the HD69830 circumstellar dust ring puts the dust at 1 ± 0.1 AU from the K0V primary (Lisse *et al.* 2007b), while recent Gemini AO measurements have shown that any warm dust in the system resides at < 2 AU from the primary (Beichman *et al.* private communication 2007). VISIR imaging constraints on the location of the warm dust in HD69830 are r < 2.4 AU (Smith, Wyatt, and Haniff, submitted).

For the HD172555 circumstellar material, scaling from the $T_{T1ejecta}$ = 340K for the hottest Deep Impact dust at 1.51 AU and $L_*$ = 1.0, we find for the smallest grains with $T_{max}$ = 335 K that $r_{dust}$ = 5.8 ± 0.6 AU. LTE for grains larger than 100 μm in radius at this distance is ~206 K. This agrees well with Wyatt *et al.* 2007b's Spitzer 24/70 μm based estimate of 4-5 AU for the location of the dust with respect to the HD172555 primary, and with the ~200K temperature of the large, cool grain dust grains from this work. The equivalent location in our solar system is at ~1.9 AU from the Sun, near the inner edge of the main asteroid belt.

A location of 5.8 AU for the circumstellar material is also consistent with the location of the innermost dust belt surrounding β Pic (6.4 AU from the star), as determined by Okamoto *et al.* (2004). From our best-fit model, the total surface area of dust that we find creating the observed mid-infrared radiation is $4 \times 10^{15} - 2 \times 10^{16}$ km$^2$. At 5.8 AU, this is equivalent to a relatively narrow and localized torus, like a dust belt, of width 0.01 AU/sin(i) (where i is the inclination of the torus to Spitzer's line of sight).

## 4. Discussion

**4.1 Physical Implications.**

**4.1.1 Three Main Reservoirs of Material: Fine dust, large cold dust, and SiO gas.**
There is $10^{19} - 10^{20}$ kg of **fine dust** in the system, corresponding to the mass of an asteroid of radius 150 - 200 km. The fine dust, dominated by $\mu$m-sized particles, is a minority component by mass but dominates much of the observed IR emission due to its large surface area and strong IR emissivity. The location of the dust, at 5.8 AU from the primary, is analogous to a locus in the inner main belt of our solar system. Because of its small size and relative transparency, we are able to determine the mineralogical composition of the fine material in some detail. High temperature, non-equilibrium amorphous silica dominates the Si and O rich fine silica dust. There is a deficit of the relatively unrefractory amorphous silicate species, and a deficit of the less refractory pyroxene versus the more refractory olivine (as compared to the primitive material found in other young systems, like the TTS and Herbig objects, Figure 7). There is a noted lack of any low-temperature organic or hydrous material. It appears probable that the fine dust material was produced from a differentiated, SiO rich body in an impulsive, high temperature event.

A component we have not had to use in our previous analyses, but find necessary to obtain a good fit to the Spitzer data (especially at the longer wavelengths) is material contained in a very **large particle dust population** (a > 100 μ m), radiating as blackbodies at ~200K, close to the local equilibrium temperature for 5.8 AU. Because it is optically thick, we cannot determine much about it mineralogical nature. While hard to characterize, most of the solid circumstellar mass, $10^{22} - 10^{23}$ kg, is in this large cold dust. It is likely that this population is a remnant from the event that created the fine dust material.

At ~$10^{22}$ kg, **SiO gas** is one of the two major reservoirs of circumstellar mass in the HD172555 system. As SiO gas is typically formed when rocky material, like olivines and pyroxenes, are vaporized in high temperature processes, it is another piece of evidence for high temperature processing. On the other hand, while the condensation temperatures for most SiO species are high, greater than 900K, the temperatures we find for the solid dust range from 200 to 450 K (Table 3). If the system were in true thermodynamic equilibrium, then there would not be any SiO gas present. We conclude that the SiO gas reservoir and the dust reservoirs are not in equilibrium, but are instead still relaxing from the recent violent event that created them - much of the material appears to be still condensing out of the vapor phase.

**4.1.2 Total Mass Scale.** Whatever processes are operative in the HD172555 system, they are acting on an important mass scale. $10^{22}$ kg of dust is about the total mass estimated to be in the solar system's main asteroid belt today. The amount of mass involved in the silica features of HD172555 is similar to the estimated amount of chondritic asteroidal material, the major reservoir of primitive silica in the present day solar system. What we are observing is not merely due to the transformation of one averaged sized asteroid, producing a dense zodiacal cloud (e.g., the HD69830 system, Lisse *et al.* 2007b) or due to the destruction of a large comet. The scale of the causal event is like that of an 'oligarch' sized planetesimal (on the order of the size of Ceres or larger). On the other hand, it is much too small to be an entire young asteroid's belt of circumstellar material – our own asteroid belt is purported to be diminished by a factor of 100 to 1000 from its original mass due to perturbations by Jupiter during the era of terrestrial planet growth; Chambers 2004).

**4.1.3 Time Scale.** The process is also operating on an interesting time scale. The presence of glassy amorphous silica-like material, produced on Earth as the kinetic product of very high temperature processing of silicaceous materials in lavas and impact craters (i.e., by quick quenching of rapidly melted rocks), and abundant SiO gas, the product of silicate vaporization, is suggestive of a quick, violent origin for the circumstellar material found around HD172555. The dominance of the SiO gas mass implies that it has not had time to recondense and settle out of the HD172555 circumstellar medium.

We estimate the time scale for SiO recondensation to occur via 2-body collisions in a circumstellar gas torus in the molecular flow regime at $\sim$ 0.1 Myr, from estimates of the collisional lifetime of zodiacal dust particles in the solar system (Burns, Lamy, and Souter 1979), assuming that when two molecules collide they stick with unit efficiency and begin condense. A similar timescale applies for sweeping the fine dust particles out of the system via radiation pressure and P-R drag (Chen *et al.* 2006). If the SiO gas is localized around a planetary body, it should recondense even more quickly. A lower limit to the timescale, $10^2$ - $10^3$ years, can be estimated from the time for lunar formation from a dense circumterrestrial torus of dust and gas created by a hypervelocity collision between the proto-Earth and the Mars-sized impacting body Theia (Pahlevan and Stevenson 2007). An absolute observational lower limit to the timescale for blowout and recondensation of 22 years can derived from the apparent lack of change between the IRAS photometry from 1983 and Spitzer fluxes from 2005 (Figure 1; Chen *et al.* 2006).

All of these processes provide strong, short upper limits (in terms of solar system evolution timescales) to the time elapsed since the event creating the observed material, explaining why such systems like HD172555 are rarely seen.

**4.2 Formation Processes in HD172555.** Given our current understanding of young stellar system evolution, the possible causes of silica and SiO gas formation in HD172555 are 4-fold: nebular shocks created by giant planet formation in the proto-planetary disk dense with gas (Desch *et al.* 2005); β-meteoroid formation and outflow, as in the Vega system (Su *et al.* 2005); massive stellar flares common to young pre-main sequence (PMS) stars (Ribas *et al.* 2005); and hypervelocity collisions in the T Tauri primordial disk due to non-Keplerian rotation and accretion.

We can quickly rule out nebular shocks as the possible source of the HD172555 silica and SiO gas, as the system is gas-poor and unable to support strong energetic shocks, having already formed gas giants (if any) at 12 Myr of age and swept out its supply of primordial gas (Chen *et al.* 2006).

We can also rule out formation by infall of 'normal' dust onto the stellar photosphere in a process similar to the creation of the β Meteoroids in the solar system (i.e., P-R drag induced infall of circumstellar dust, evaporation of the dust as it nears the stellar surface, and radiation pressure blowout of the micron-sized remnants; McDonnell and Dohnanyi 1978). The photo-evaporation mechanism would create copious amounts of SiO gas and smelt large amounts of silicaceous material, separating SiO from MgO, FeO, CaO, etc., into ~1 µm particles at temperatures ~1000 - 1500 K, and spread them over 10-100 AU as they are ejected from the system (e.g. the Vega system morphology, Su *et al.* 2005). By contrast, the highest temperatures measured in our study for the (smallest) HD172555 dust particles is ~300K, the dust is localized at ~5.8 AU, and it is commingled with a population of large dust in radiative thermal equilibrium with temperatures ~200K.

Finally, we can rule out a stellar flare driven silica formation scenario. As a late type A-star, HD172555 should be relatively x-ray flare quiet, and indeed has a low measured fractional x-ray luminosity $L_x/L_{bol}$ of $2\times10^{-6}$ (Chen *et al.* 2006). Also, the size of a flare required to melt and vaporize $10^{22}$ kg of SiO gas (roughly equivalent to a 1000 km radius asteroid) is enormous (at $\Delta H_{vap}$ ~ 50 MJ/kg, Gail 2002), $10^{22}$ kg of rock requires $10^{30}$ J total vaporization energy, and a 1000 km radius asteroid at 5.8 AU would require a total flare energy of $4 \times 10^{43}$ J emitted into $4\pi$ steradians. Even for flares lasting years ($10^7$-$10^8$ sec), we require flares of $L_x \geq 10^{35}$ J sec$^{-1}$, prohibitive by many orders of magnitude versus the maximal observed $L_x$ values of $10^{23}$ J sec$^{-1}$ for Herbig Ae stars and main sequence A-stars (Skinner *et al.* 2004).

We are thus left with a hypervelocity collision as the likely cause of the observed circumstellar material. The fine silica dust, SiO gas, and a large amount of macroscopic rubble were created in recent large hypervelocity ($v_{impact} > 10$ km sec$^{-1}$) impact(s), via high temperature processes operating at or near the surface of contact of the impacting bodies. In these impacts, multiple processes occur: the kinetic energy of the incoming bolide is transformed directly into heating and vaporization of rock into SiO gas at the surface of contact, a large amount of material is also spalled off in the liquid phase, to soldify in-flight as tektite-like material, while at the edges of the affected region, shock waves propagate and excavate solid material in rubble form, producing a population of macroscopic, large dust particles (Melosh 1989). The impact disrupts the colliding body(s), while releasing into circumstellar orbit a massive amount of material. The results of this collision are being seen soon after the impact, so there has been little time for the silica dust and SiO gas to relax to the more thermodynamically stable olivine and pyroxene materials, or to be superceded by the products of subsequent grinding of the macroscopic fragments.

Hypervelocity impacts with enough mass and specific kinetic energy ($V_{impact} > 10$ km sec$^{-1}$; Grieve and Cintala 1992, Cintala and Grieve 1998) could occur in events similar to the lunar formation event between large differentiated planetesimals. They could also occur due to the stirring up or by destabilization of an asteroid belt during oligarch formation, or by planetary migration through a local dynamical resonance, although these are somewhat more complicated scenarios, requiring a large number of similar impacts. Another possibility for achieving a relative impact velocity of > 10 km sec$^{-1}$ is gravitational focusing of a bolide onto a large body. The relative impact velocity achieved will then be roughly the escape velocity of the large body. A simple calculation for a large body made of terrestrial planet material shows that it would have to mass about as much or more than the Earth ($V_{escape, Earth}$ = 11 km sec$^{-1}$).

Observationally, the hypervelocity collision mechanism appears highly plausible for the HD172555 system, as (a) the closely analogous β Pic system shows evidence for large clumped regions of enhanced dust due to collisions (Telesco *et al.* 2005), three belts of enhanced IR emission from dust created by planetary formation (Okamato *et al.* 2004), and evidence of the existence of a planet at ~8 AU (Lagrange *et al.* 2009);  (b) the amount of dusty material and gas we detect in a narrow torus at 5.8 AU is equivalent to at least a 1000 km radius asteroid of 2.5 g cm$^{-3}$ density, the size of oligarchs and proto-planets; and (c) the 10-100 Myr time of formation of the Moon in a massive impact on the proto-Earth and the stripping of Mercury's surface layers, is consistent with the estimated ~12 Myr age of HD172555 and (d) after a small bolide of ~370 kg mass impacted onto a comet nucleus at 10.2 km sec$^{-1}$, the recent Deep Impact hypervelocity experiment produced analogous materials to what we have found for HD17255 : re-frozen silicaceous droplets, i.e. fine dust, volatilized cometary material, i.e. SiO gas, and excavated, unprocessed bulk material, i.e. large cool dust grains (A'Hearn *et al.* 2005, Melosh 2006, Richardson *et al.* 2007).

One consequence of the hypervelocity impact model is that the silica dust and SiO gas in the HD172555 system should be relatively transient in nature, clearing on timescales of < 1 Myr (Sec 4.1), somewhat in contrast to the predictions of Wyatt et al (2007a). Another consequence is that we can use the presence of strong mid-IR 8-9 µm silica and SiO gas features to locate massive hypervelocity events in non-primordial exo-solar systems. There is some indication, for example, that the ~100 Myr old star HD23514, reported to have massive amounts of dust created by binary collisions, also demonstrates strong ~9 µm emission spectral features (Rhee *et al.* 2008).

**4.2 Silica in other Stellar Systems.** We discuss here, for the sake of completeness, the nature of other stellar objects known to exhibit the unusual silica infrared emission feature. Comparison to other astrophysically similar spectra of comparable quality (Figure 2) verifies the unusual nature of HD172555. Silica has not been reported in the ISM nor is it found in the majority of dusty exo-systems. Some of these other systems, like BD+20 307 (Song *et al.* 2005), HD69830 (fragmentation of the equivalent of a 30 km radius P or D asteroid at 2 Gyr, Beichman *et al.* 2005, Lisse *et al.* 2007b) and HD113766 (fragmentation of the equivalent of > 300 km S-type asteroid, Lisse *et al.* 2008) have been implicated as having large amounts of silicate-rich dust due to low velocity collisional processes that rubbleizes bodies, without creating much alteration of their mineralogy or chemistry. Evidence for silica based emission has also not been reported in cometary dust. This is not surprising, as silica is not the low energy thermodynamic equilibrium state of Si, O, Mg, and Fe (Lewis 1989). Si-O gas masers do exist around late-type oxygen rich AGB stars, but these require a large amount of free Si-O gas created through stellar mass shedding, driven by the local high stellar photon flux, to produce a population inversion and 8 - 9 µm emission. A handful of dense cloud cores have been reported to have Si-O masers under similar conditions.

Out of hundreds of mid-infrared spectra compiled by E. Furlan (2008, private communication), we have found five other spectra which exhibit strong silica features at ~ 9 µm: four TTS (FN Tau, CY Tau, V410 Anon 13, Hen3-600, published by Sargent *et al.* 2006) plus the 100 Myr old HD23514 (Rhee *et al.* 2008). The infrared emission spectrum of HD172555 is in fact very similar to the FN Tau TTS spectrum of Sargent *et al.* (Figure 8). The close similarity of the FN Tau and HD172555 spectra have in fact lead us to consider the misidentification of HD172555 as an early MS star. We can rule out the misidentification of HD172555 as a late type HAe or early type TTS, however, due to its advanced age by association with β Pic, its low x-ray luminosity ($L_x/L_{bol} < 2\times10^{-5}$; de la Reza and Pinzon 2004, Schröder and Schmitt 2007), and its low disk gas content (Chen *et al.* 2006 and this work).

While HD172555 is clearly not a TTS, could the same mechanisms be forming silica in the two cases? The physical situation around a TTS is much different than for circumstellar material in a debris disk around a young MS star like HD172555. Massive amounts (~0.01 $M_\odot$) of finely divided primordial material are located around a PMS star in an optically thick circumstellar disk many AUs in radius. This material is continually acted upon and altered over the $10^6$ - $10^7$ years of existence of the TTS phase. The most likely causes of silica formation in T Tauri systems are nebular shocks, massive stellar flares, and hypervelocity collisions (Sec 4.2). While we expect shocks and stellar flares in the gas-rich, x-ray active T Tauri systems, we do note that collisions, like the aftermath of the one we are observing in HD172555, should be very frequent in a thick massive T Tauri disk as the rate of collision goes at least as ~ disk density squared, and can even be enhanced due to gravitational focusing as giant planet formation proceeds though collisional accretion in the first few Myr.

**4.3 Solar System Analogues.** There is abundant inferential evidence for massive collisions in the early solar system (Hartmann and Vail 1986): Mercury's high density; Venus' retrograde spin; Earth's Moon; Mars' North/South hemispherical cratering anisotropy; Vesta's igneous origin (Drake 2001); brecciation in meteorites (Bischoff *et al.* 2006); Uranus' spin axis located near the plane of the ecliptic. Local geological evidence for widespread impact melting includes tektites found on Earth and glass beads found in lunar soils (chemically distinguished from volcanic glasses also found in the soil).

Solar system bodies without significant tectonic reworking, weathering or erosion of their surfaces, like the Moon, Mercury, and Vesta, have the best evidence for giant impacts during their early growth stages. Of the three bodies, the case of Mercury is the most likely analogue of the occurrences in the HD172555 system. In the lunar formation event, the incoming bolide Theia had to have impacted at a relative speed about equal to the escape velocity for the Earth-moon system ($v_{escape}$ = 12 km sec$^{-1}$), so that the bulk of the excavated material either fell back to the Earth or remained bound in a circumterrestrial torus or disk rather than going into solar orbit (Canup 2004, 2008). Massive amounts of iron- and volatile poor rock was vaporized, melted, and extracted from the upper layers of the proto-Earth, and a magma ocean was induced on the Earth's surface. However, while physically analogous to the event we expect formed the HD172555 materials, an event channeling most of the excavated material into a small, dense disk with dimensions on the order of the Earth-Moon distance would be hard to detect by Spitzer. In fact, the huge surface area of the dust creating the strong HD172555 mid-infrared signature, $\sim 10^{16}$ km$^2$, implies an equivalent disk radius of ~0.4 AU (6 x 10$^7$ km), more than two orders of magnitude larger than the Earth-Moon distance of ~4x 10$^5$ km, and far outside the Earth's Hill sphere.

Little is currently known about the **origins of Vesta**, other than it was formed in the first few million years of the solar system, its surface is highly variegated and heavily scarred by massive cratering, and that it consists of a highly basaltic igneous mineralogy that has been thoroughly heated and differentiated. Spectra of Vesta's surface (Vernazza *et al.* 2005), and Vesta-like meteorites (the so called HED meteorites, Usui and McSween 2007) indicate the presence of silica, but as a minority component at the ~10% level. Thus while we note its possible connections with the case of HD172555, as local evidence for an asterodial source of silica dust and SiO gas, we do not discuss it further, other than to note the following plausibility argument for its early participation in a hypervelocity (> 10 km sec$^{-1}$) impact event. Keplerian orbital velocities of main belt asteroids (MBAs) in the solar system are on the order of $v_{Kepler} \sim \sqrt{(G*M_\odot/2.6\ AU)}$ = 18.5 km sec$^{-1}$. Along with $v_{Kepler}$, the dispersion in orbital inclinations of the MBA population is a critical defining characteristic for the velocity of interaction $v_{inter}$ between two objects, as most collisions and velocity mismatches occur mainly in the direction perpendicular to the general Keplerian flow (Bottke *et al.* 1994). E.g., in today's solar system, the ± 15 deg range of inclinations versus the ecliptic in today's main belt leads to an average $v_{inter}$ of ~5 km sec$^{-1}$. The young main asteroid belt at a few Mya age is predicted to have been frequently dynamically excited as planetary accretion progressed and oligarch formation occurred, leading to much larger inclination dispersions (up to ± 45 deg), and much more energetic collisions than in the present day (Chambers 2004, Kenyon and Bromley 2004, 2006). Collision rates were also orders of magnitude higher in the much denser primordial belt.

The dust and particles in the HD172555 system we have found at 5.8 AU should be moving at Keplerian orbital velocities with respect to the HD172555 primary of roughly $v_{Kepler} = \sqrt{(G*M_*/r_*)} = \sqrt{(G*2M_\odot/5.8AU)} = 17.5$ km sec$^{-1}$, similar to the velocities of main belt asteroids (MBAs) in our solar system. Also like the young main belt, inclination dispersions expected of young systems such as HD172555 should be dynamically excited and much more 'puffed up', with a large inclination dispersion and a correspondingly much greater chance of undergoing a hypervelocity interaction with $v_{inter} > 10$ km sec$^{-1}$.

The details of a **crustal stripping event** are fully consistent with our mineralogical findings, and can easily create the amount of dust we observe in HD172555. A giant impact event akin to the one proposed to have stripped Mercury of its surface layers of light crustal rock, throwing ejected material into a circumstellar torus centered on the impacted body's orbit, is expected to have occurred at high relative velocities of $v_{inter} \geq$ 20 km sec$^{-1}$ (D. Stevenson, private communication 2008) and to have lasted long enough (~$10^5$ yrs, Sec 4.1) to be detected by Spitzer. Modeling the Mercury-stripping event, Benz *et al.* (1988) found that proto-Mercury was ~2.25 times more massive than at present, before suffering a hypervelocity impact at 20 – 35 km sec$^{-1}$ by a body 0.16 as massive as today's Mercury. In their models, the impactor became totally vaporized, indicating that temperature and pressure regimes leading to silica and SiO gas formation were reached. The 1.4 $M_{Mercury}$ of ejected material is dominated, however, by material from the differentiated surface layers of Mercury, creating the depletion of Mg, Fe, S, etc. vs. solar that we see in the HD172555 material. Modern predictions of the short term recondensation products from the Mercury stripping event, following models similar to the equilibrium study of Fegley and Cameron (1987) would be very interesting to compare to what we find in the Spitzer spectra of HD172555.

## 5. Conclusions

The circumstellar dust detected by IR emission around the close β Pic analogue star HD172555 is very peculiar. We know of only one other post-T Tauri system and three other T Tauri systems, out of hundreds of observed stellar systems, that show this kind of infrared spectrum.

If our inferred composition is correct, the dust mineralogy is very unusual, composed of three fractions: a fine dust component, a large cold dust component, and a reservoir of SiO gas. The **fine, micron sized dust** dominates the emitting surface area of circumstellar material and is composed of exclusively refractory materials, with approximately three-quarters of the Si atoms in high temperature, non-equilibrium amorphous silica species. The remaining fine dust fractions are rich in refractory crystalline olivines and to a lesser extent crystalline pyroxenes. There are almost no amorphous silicates present in the material. No water or water ice, carbonates, phyllosilicates or amorphous carbon appear to be present. A comet-like body can be ruled out as the parent source on compositional grounds. We find a steep, small particle dominated best-fit model size distribution for the observed dust of $dn/da = a^{-3.95\pm0.10}$. This steep a size distribution, with abundant micron-sized particles, argues for a fresh source of material within the last 0.1 Myr (Chen *et al.* 2006).

A broad emission feature centered at ~8 μm due to the **SiO gas** fundamental is also found, as is evidence for the rise in the emission spectrum towards the 4 μm first overtone of SiO. Roughly $10^{22}$ kg of SiO gas is present in the system, dominating the reservoir of detected Si and O atoms. The circumstellar material does not appear to be in chemical equilibrium, and we expect the SiO gas to recondense within the next 0.1 Myr.

A significant population of large dust particles radiating as blackbodies at ~200 K (LTE for 5.8 AU for HD172555 is 206 K) is also found commingled with the small dust producing the silica and silicate emission features. The composition of this material is indeterminate. The temperature of the large cold dust, 200K, as well as the hottest fine dust particles detected, 305 K, places the location of the dust with respect to the star at 5.8 ± 0.6 AU (at the rough equivalent of 1.9 ± 0.15 AU from our Sun, in the inner main asteroid belt), within the terrestrial planet formation region but at the outer edge of any possible terrestrial habitability zone for HD172555. A similar belt of material has been reported at 6.4 AU from the close analogue star β Pic. The lower mass limit for the amount of large dust and gas, ~$10^{22}$ kg, is equivalent to a ~1000 km radius, 2.5 g cm$^{-3}$ asteroid. This mass range equals the largest sized asteroids currently found in the solar system's main asteroid belt, and the upper end of the range includes the smaller sized oligarchs expected to form during terrestrial planet building.

The most likely cause of the unusual mineralogy is a hypervelocity (> 10 km sec$^{-1}$) impact between two rocky planetesimals, similar to the ones which created the Moon, melted and differentiated Vesta, and stripped the surface crustal material off of Mercury's surface in the Solar System, with the latter scenario being the most understood and likely. We hypothesize that the large, blackbody-like dust grains are the macroscopic fragments left over from the hypervelocity collision, while the small grains are the material recondensed (and recondensing) from the melt and SiO vapor created by the impact.

## 6. Acknowledgements

This paper was based on observations taken with the NASA *Spitzer* Space Telescope, operated by JPL/CalTech. C. M. Lisse gratefully acknowledges support for performing the modeling described herein from JPL contract 1274485 and the APL Janney Fellowship program**.** The authors would also like to thank W. Bottke, D. Ebel, H. Levinson, D. Nesvorny, A. Roberge, B. Sargent, G. Sloan, D. Stevenson, and D. Watson for many valuable discussions concerning this work, and M. Grady and C. Smith of the Natural History Museum, London for chemical analysis of the Bediasite tektite sample.

# 7. References

Ardilla, D. *et al.* 2008. "A Search for Debris Disks in the Coeval b Pictoris Moving Group", *Astrophys. J.* (in preparation) (also ACS/HRC/WFC program abstract # 10487)

Beichman, C. A., *et al.* 2005, *ApJ* **626**, 1061

Benz, W., Slattery, W.L., and Cameron, A.G.W 1988,*Icarus* **74**, 516

Bischoff, A. *et al.*, 2006. In Meteorites and the Early Solar System II, D. S. Lauretta and H. Y. McSween Jr. (eds.), University of Arizona Press, Tucson, 943 pp., p.679-712

Bottke, W.F., *et al.* 1994, *Icarus* **157**, 205

Bradley, J.P., *et al.* 1999, *Science* **285**, 1716

Bradley, J. *et al.* 2005, *Science* **307**, 244

Brearley and Jones, 1998 In: Planetary Materials, Reviews in Mineralogy 36)

Burns, J.A., Lamy, P.L., and Soter, S. 1979, *Icarus* **40**, 1

Canup, R.M., 2004, *Ann. Rev. A&A* **42**, 441

Canup, R.M., 2008, *Icarus* **196**, 518

Cardelli, J. A., Clayton, G. C., & Mathis, J. S. 1989, *ApJ* **345**, 245

Chambers, J.E., 2004, *Earth and Plan Sci Lett* **223**, 241

Chen, C. H., *et al.* 2006, *ApJS* **166**, 351

Chen. C, *et al.* 2007, *ApJ* **666**, 466

Cintala, M.J. and Grieve, R.A., 1998, *Meteoritics and Planetary Science* **33**, 889

Cohen, M., & Davies, J. K. 1995, *MNRAS* **276**, 715

Crovisier, J. 2002. "Constants for Molecules of Astrophysical Interest in The Gas Phase: Photodissociation, Microwave and Infrared Spectra, V4.2"  http://www.lesia.obspm.fr/~crovisier/basemole/intro42.ascii

de la Reza, R. and Pinzon, G. 2004, *AJ* **128**, 1812

Desch, S.J., *et al.* 2005 and references therein. "Heating of Chondritic Materials in Solar Nebula Shocks", in *Chondrites and the Protoplanetary Disk,* ASP Conference Series **341**, Proceedings of a workshop held 8-11 November 2004 in Kaua'i, Hawai'i. Edited by A.N. Krot, E.R.D. Scott, and B. Reipurth. San Francisco: Astronomical Society of the Pacific, 84

Dohnanyi, J. W. 1969, *JGR* **74**, 2531

Draine, B.T. and Li, A., 2007, *ApJ* **657**, 810

Drake, M.J. 2001, *Meteoritics and Planetary Science* **36**, 501

Drira, I., *et al.* 1997, *A&A* **319**, 720

Fegley, B. and Cameron, A.G.W. 1987. *Earth and Planetary Science Letters,* **82**, 207

Feigelson, E.D., *et al.* 2006, *AJ* **131,** 1730

Gail, H.-P., 2002, In *Astromineralogy*, *Lecture Notes in Physics* **609**, ed. T.K. Henning, pp. 55-120

Grady, C.A., *et al.* 2001, *AJ* **122**, 3396

Grady, C.A., *et al.* 2005, *ApJ.* 620, 470

Gray, R.O., *et al.* 2006, *AJ* **132**, 161

Grieve, R.A. and Cintala, M.J., 1992, *Meteoritics* **27**, 526

Hartmann, W.K., and Vail. S.M., 1986. "Giant Impactors - Plausible Sizes and Populations", in *Origin of the Moon*, Proceedings of the Conference, Kona, HI, October 13-16, 1984 (A86-46974 22-91). Edited by W.K. Hartmann, R.J. Phillips, and G.J. Taylor. Houston, TX, Lunar and Planetary Institute, 551

Houk, N. and Cowley, A.P. 1975. Michigan Catalogue of Two-dimensional Spectral Types for the HD Stars, Ann Arbor: University of Michigan, Department of Astronomy, 1975

Kenyon, S. J., and Bromley, B. C. 2004, *ApJ Lett.* **602**, L133

Kenyon, S. J., and Bromley, B. C. 2006, *AJ* **131**, 1837

Koeberl, C., 1988, *Meteoritics* **23**, 161

Kudo, T., *et al.* 2008, *ApJ* **673**, L67

Lacy, C. H., 1977, *ApJS* **34**, 479

Lagrange, A.-M. *et al.* 2009, *A&A* **493**, L21

Lewis, J.S., 1989, "Cosmochemistry", in *The Formation and Evolution of Planetary Systems*, *Proceedings of the Meeting, Baltimore, MD, May 9-11, 1988 (A90-31251 12-90).* Cambridge and New York, Cambridge University Press, 309

Lisse, C. M., *et al.* 2006, *Science* **313**, 635

Lisse, C.M., *et al.* 2007a, *Icarus* **187**, 69

Lisse, C. M., *et al.* 2007b, *ApJ* **658**, 584

Lisse, C.M., *et al.*, 2008, *ApJ* **673**, 1106

McDonnell J.A.M. and Dohnanyi J.S. 1978. Beta-Meteoroids. In *Cosmic Dust,* ed. J. A.M. McDonnell, Wiley, New York, pp. 527–605

Melosh, H.J., *Impact Cratering: A Geologic Process.* Oxford Univ. Press: New York (1989).

Melosh, H.J., *et al.* 2006. *Lunar Planet. Sci.* **37**, 1165.

Moór, A., *et al.* 2006, *ApJ* **644**, 525

Morales, F.Y., *et al.* 2009, *ApJ*, in press

Okamoto, Y. K., *et al.* 2004, *Nature* **431**, 660

Pahlevan , K., and Stevenson, D.J., 2007, Earth and Planetary Science Letters **262**, 438

Reach, W.T., *et al.* 2003, *Icarus* **164**, 384

Reach, W.T., Lisse, C.M., von Hippel, T. and Mullally, F., 2008, *ApJ* **693**, 697

Rhee, J.H., Song, I., and Zuckerman, B. 2008, *ApJ* **675***, 777*

Ribas, I. *et al.* 2005, *ApJ* **622**, 680

Richardson, J.E, *et al.*, 2007, *Icarus* **190**, 357

Roberge, A., *et al.* 2006, *Nature* **441**, 724

Sargent, B., *et al.* 2006, *ApJ* **645**, 395

Schröder, C. and Schmitt, J. H. M. M. 2007. *A&A* **475**, 677

Schutz, O., Meeus, G., & Sterzik, M. F. 2005, *A&A* **431**, 175

Skinner, S.L., *et al.* 2004, *ApJ* **614**, 221

Smith, R., Wyatt, M.C., and Haniff,  C.A. 2009. *A&A*, submitted

Song, I., *et al.*,  2005, *Nature* **436**, 363

Su, K.Y.L., *et al.* 2005, *ApJ* **628**, 487

Su, K.Y.L., *et al.* 2006, *ApJ* **653**, 675

Telesco, C. M., *et al.* 2005, *Nature* **433**, 133

Tissandier, L., Libourel, G., and Robert, F. 2002. *Meteoritics & Planetary Science* **37**, 1377

Usui, T. and McSween, Jr., H. Y., 2007, *Meteoritics and Planetary Science* 42, 255

Vernazza, P. *et al.* 2005, *A&A* **436**, 1113

Warren, P.H. 2008, *Geochimica et Cosmochimica Acta* **72**, 3562

Watson, D., *et al.* 2004, *ApJS* **154**, 391

Wetherill, G.W. 1990. *Annu. Rev. Earth Planet Sci.* **18**, 205

Wyatt, M. C., *et al.* 2007a, *ApJ* **658**, 572

Wyatt, M. C., *et al.* 2007b, *ApJ* **663**, 365

Yin, Q., *et al.* 2002, *Nature* **418**, 949

Zuckerman, B., & Song, I. 2004, *ARA&A* **42**, 685

# 8. Tables

## Table 1. Properties of Star HD172555 (HR 7012)

| Name | Sp Type | $T_*$ (°K) | $M_\star$ ($M_\odot$) | $R_\star$ ($R_\odot$) | $L_\star$ ($L_\odot$) | d (pc) | Age (Myr) | $f_{IR}/f_{bol}$ | $r_{dust}$ (AU) | $v_{dust}$ (km sec$^{-1}$) |
|---|---|---|---|---|---|---|---|---|---|---|
| HD172555A | A5 IV/V | 8000 | 2.0 | 1.6 | 9.5 | 29.2 | 12 | $5 \times 10^{-4}$ | 5.8 | 17.5 |

- Data from Wyatt *et al.* 2007b.  - If dust is at 5.8 AU, and $M^* = 2M_\odot$, then $v = \sqrt{(GM/r)} = \sqrt{(G*2M_\odot/5.8AU)} = \sqrt{(GM_\odot/2.9AU)}$

## Table 2. Composition of the Best-Fit Model[a] to the SST IRS HD172555 Spectrum

| Species | Weighted[b] Surface Area | Density (g cm$^{-3}$) | M.W. | $N_{moles}$[c] (relative) | Model $T_{max}$[d] (°K) | Model $\chi^2_\nu$ if not included |
|---|---|---|---|---|---|---|
| **Detections** | | | | | | |
| *Olivines* | | | | | | |
| AmorphSilicate/Olivine (MgFeSiO$_4$) | 0.09 | 3.6 | 172 | 0.19 | 305 | 2.32 |
| ForsteriteKoike (Mg$_2$SiO$_4$) | 0.17 | 3.2 | 140 | 0.39 | 305 | 1.69 |
| Fayalite (Fe$_2$SiO$_4$) | 0.09 | 4.3 | 204 | 0.19 | 305 | 1.69 |
| *Pyroxenes* | | | | | | |
| FerroSilite (Fe$_2$Si$_2$O$_6$) | 0.10 | 4.0 | 264 | 0.14 | 305 | 1.44 |
| Diopside (CaMgSi$_2$O$_6$) | 0.05 | 3.3 | 216 | 0.075 | 305 | 1.11 |
| OrthoEnstatite (Mg$_2$Si$_2$O$_6$) | 0.06 | 3.2 | 200 | 0.09 | 305 | 1.23 |
| *Silica, SiO* | | | | | | |
| Tektite (Bediasite, 66% SiO$_2$, 14% Al$_2$O$_3$, 7% MgO, 6% CaO, 4% FeO, 2% K$_2$O, 1.5% Na$_2$O, Koeberl 1988) | 0.35 | 2.6 | 62. | 1.47 | 305 | 11.2 |
| Obsidian (75% SiO$_2$, 18% MgO, 6% Fe$_3$O$_4$) | 0.18 | 2.6 | 66 | 0.71 | 305 | 1.34 |
| SiO Gas (100% SiO) | 0.20 | N/A | 44 | 4-8 | 305 | 1.17 |
| *Metal Sulfides* | | | | | | |
| Pyrrhotite (as Mg$_{10}$Fe$_{90}$S) | 0.16 | 4.5 | 84 | 0.86 | 305 | 6.63 |
| *Organics* | | | | | | |
| Amorph Carbon (C) | 0.09 | 2.5 | 12 | 1.9 | 450 | 1.32 |
| **Upper Limits and Non-Detections** | | | | | | |
| AmorphSilicate/Pyroxene (MgFeSi$_2$O$_6$) | $\leq 0.01$ | 3.5 | 232 | $\leq 0.015$ | 305 | 1.03 |
| *Water* | | | | | | |
| Water Gas (H$_2$O) | $\leq 0.01$ | 1.0 | 18 | $\leq 0.056$ | 200 | 1.03 |
| Water Ice (H$_2$O) | $\leq 0.01$ | 1.0 | 18 | $\leq 0.056$ | 200 | 1.03 |
| *Carbonates* | | | | | | |
| Magnesite (MgCO$_3$) | $\leq 0.01$ | 3.1 | 84 | $\leq 0.037$ | 305 | 1.03 |
| Siderite (FeCO$_3$) | $\leq 0.01$ | 3.9 | 116 | $\leq 0.034$ | 305 | 1.03 |
| *PAHs* | | | | | | |
| PAH (C$_{10}$H$_{14}$) | $\leq 0.01$ | 1.0 | <178> | $\leq 0.011$ | N/A | 1.03 |

(a) - Best-fit model $\chi^2_\nu = 1.03$ with power law particle size distribution $dn/da \sim a^{-4.0}$, 5 - 33 μm range of fit. 95% C.L. hás $\chi^2_\nu = 1.07$
(b) - Weight of the emissivity spectrum of each dust species required to match the HD172555 emissivity spectrum. Weight of 200K BB = 1.2
(c) - $N_{moles}(i) \sim$ Density(i)/Molecular Weight(i) * Normalized Surface Area (i). Errors are $\pm 10\%$ (1σ). $N_{Gas\,moles}$ depends on the line emission calculation of Sec. 3.2
(d) - All temperatures are $\pm 10K$ (1σ). Best fit blackbody function to 5 - 33 μm IRS spectrum hás $T_{bb} = 335$ K.

**Table 3. Refractory Dust Atomic Abundances for Solar for Objects Observed by ISO/*Spitzer*[a]**

| Object | H | C | O | Si[b] | Mg | Fe | S | Ca | Al |
|---|---|---|---|---|---|---|---|---|---|
| Tempel 1 ejecta | 3.8e-4 | 0.052 | 0.46 | 1.0 | 0.82 | 0.79 | 0.61 | 0.84 | 1.0 |
| Hale-Bopp coma | 5.0e-05 | 0.13 | 0.23 | 1.0 | 1.1 | 0.97 | 0.63 | 0.45 | 1.4 |
| HD100546 (Be9V) | 1.1e-4 | 0.35 | 0.26 | 1.0 | 0.88 | 0.76 | 0.63 | 0.00 | 1.2 |
| HD69830 | 8.8e-06 | 0.088 | 0.18 | 1.0 | 1.1 | 0.43 | 0.00 | 0.93 | 0.0 |
| HD113766 (F3/F5) | 3.0e-05 | 0.076 | 0.18 | 1.0 | 0.77 | 1.09 | 1.73 | 0.82 | 0.96 |
| HD172555 (A5) [fine dust] | 0.0 | 0.064 | 0.13 | 1.0 | 0.45 | 0.59 | 0.55 | 0.34 | 1.64 |

[a] - Abundance estimates have $2\sigma$ errors of $\pm 20\%$.
[b] - All abundances are with respect to solar, with the Si abundance assumed to be = 1.0 for normalization purposes.

**Table 4. Derived Total Masses (in beam) for the Objects Observed by ISO/*Spitzer* and Selected Relevant Solar System Objects**

| Object | Observer Distance[1] (pc/AU) | Mean Temp[2] (K) | Equiv Radius[3] (km) | 19 um Flux[4] (Jy) | Approximate Mass[5] (kg) |
|---|---|---|---|---|---|
| Earth | --- | 282 | 6380 | | $6 \times 10^{24}$ |
| Mars | 1.5 AU | 228 | 3400 | | $6 \times 10^{23}$ |
| Mercury | 0.5-0.7 AU | 400 | 2400 | | $3 \times 10^{23}$ |
| HD113766 (F3/F5) | 130.9 pc | 440 | 300 - 3000 | 1.85 | $7 \times 10^{20}$ - $3 \times 10^{23}$ |
| Moon | 0.0026 AU | 282 | 1740 | | $7 \times 10^{22}$ |
| HD172555 (A5) | 29 pc | 335 | 1000-2000 | | $10^{22}$ - $10^{23}$ |
| HD100546 (Be9V) | 103.4 pc | 250/135 | $\geq 910$ | 203 | $\geq 1 \times 10^{22}$ |
| Pluto | 40 AU | 45 | 1180 | | $1 \times 10^{22}$ |
| Asteroid Belt | 0.1 - 5.0 AU | Variable | | | $3 \times 10^{21}$ |
| HD69830 | 12.6 pc | 340 | 30 - 60 | 0.11 | $3 \times 10^{17}$ - $2 \times 10^{18}$ |
| Zody Cloud | 0.1 - 4.0 AU | 260 | | | $4 \times 10^{16}$ |
| Asteroid | 0.1 - 5.0 AU | Variable | 1 - 500 | | $10^{13}$ - $10^{21}$ |
| Comet nucleus | 0.1 - 10 AU | Variable | 0.1-50 | | $10^{12}$ - $10^{15}$ |
| Hale-Bopp coma | 3.0 AU | 200 | | 144 | $2 \times 10^{9}$ |
| Tempel 1 ejecta | 1.51 AU | 340 | | 3.8 | $1 \times 10^{6}$ |

(1) - Distance from Observer to Object.
(2) - Mean temperature of thermally emitting surface.
(3) - Equivalent radius of solid body of 2.5 g cm$^{-3}$.
(4) - System or disk averaged flux.
(5) - Lower limits are conservative, assuming maximum particle size of 1000 µm, and ignoring optical thickness effects. For HD172555, we have u=included the mass of SiO gas in the estimate. Upper limits assume a maximum particle size of 10m radius.

# 9. Figures

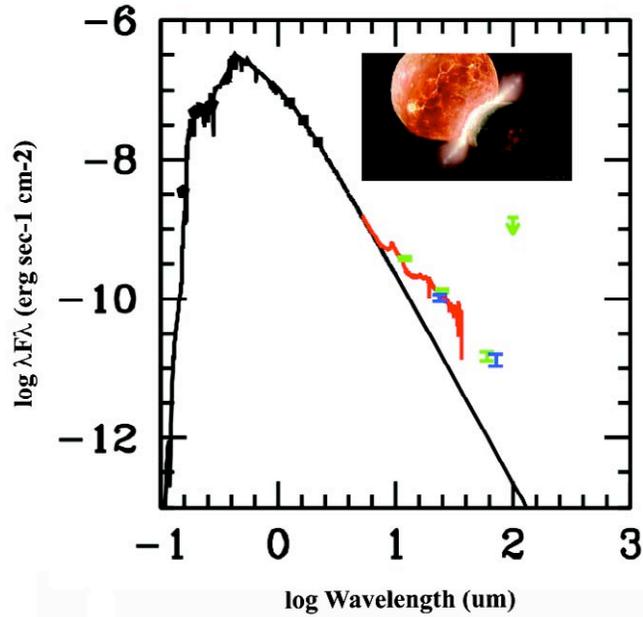

**Figure 1** - *Left :* SED for HD172555 showing the BVR/2MASS determined photospheric emission (black) and the large IR excess detected by both IRAS (green) and Spitzer MIPS (blue) and Spitzer IRS (red). *Right :* Color-magnitude comparison of nearby IRAS excess stars. The red circle marks the similar locations of HD172555 and β Pic on the diagram. (After Moór *et al.* 2006).

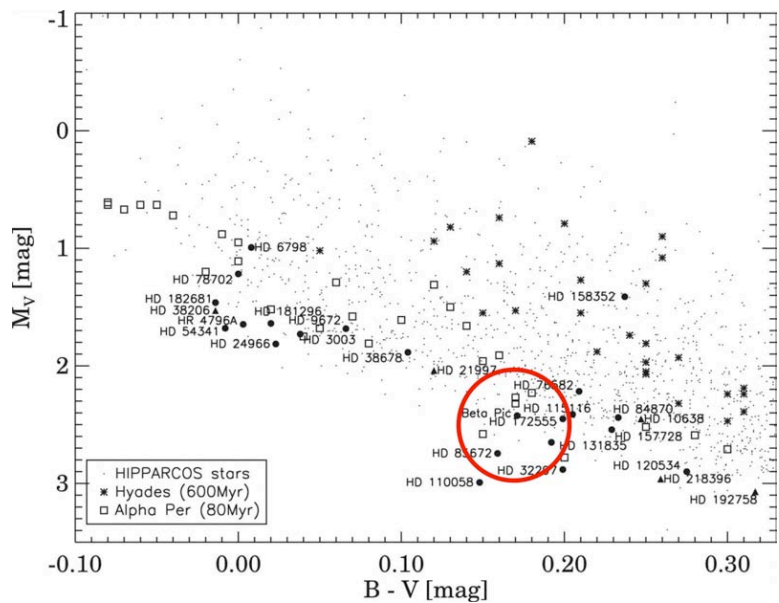

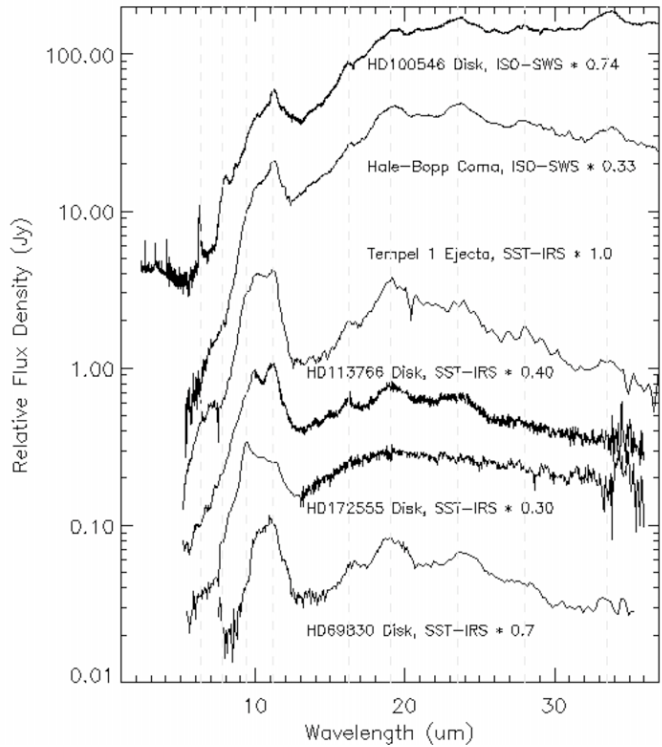

**Figure 2** - Comparison of the mid-IR spectra of HD172555 with the spectra of dust from: two comets (Hale-Bopp and Tempel 1); a young, organic rich Herbig A0 star building a giant planet (HD100546); a young F5 star building a terrestrial planet (HD113766); and a mature main sequence star with a dense zodiacal cloud (HD69830). The unusual nature of the 7 - 13 μm emission from HD172555 is readily apparent.

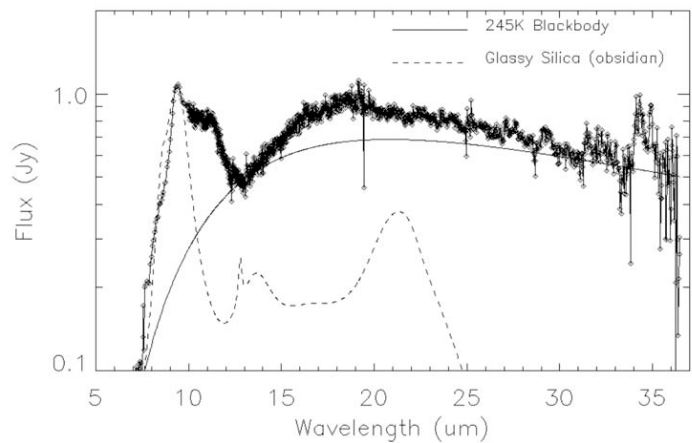

**Figure 3** - Comparison of the Spitzer IRS HD172555 excess spectrum to a blackbody model and the spectrum of an ~ 1 μm powder obsidian sample. The 9.3 μm Si-O silicate stretch feature is clearly identified above the blackbody continuum. The excess shoulder at 7.5 - 8.2 μm due to Si-O gas emission can also be seen. Error bars are 2σ.

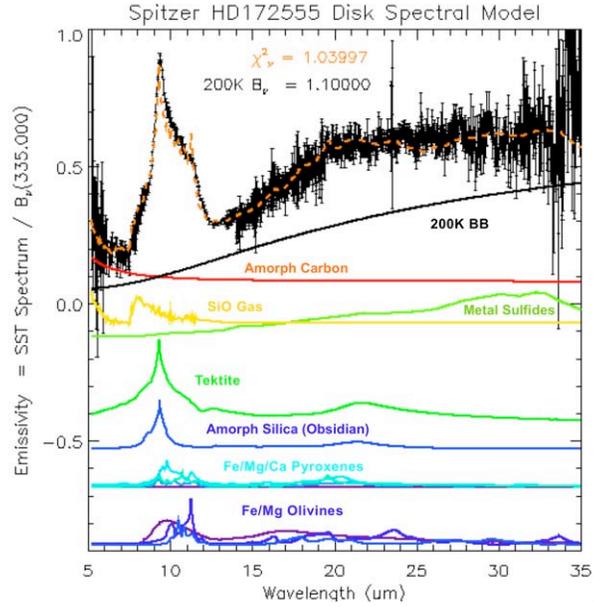

**Figure 4** - Spitzer IRS emissivity spectrum of the HD172555 circumstellar excess, and best-fit compositional model. Relative abundances for the best-fit model are given in Table 1. *Top*- Best fit model using small, solid, optically thin dust grains, SiO gas, and a population of large, cold dust grains (200 K blackbody). The relative contribution of each species to the total observed flux is given by the amplitude of each emissivity spectrum. A 335K blackbody temperature dependence has been removed from the as-observed flux to create the emissivity spectrum. *Bottom* - Residuals of the best-fit model to the Spitzer IRS emissivity spectrum. All of the usual **silicate** species have been accounted for and removed. The contributions of the important unusual species SiO gas (yellow), amorphous, glassy silica ('Tektite', green and 'Obsidian', blue) can clearly be seen, as can the emissivity contributions of the more typical species amorphous carbon (red) an metal sulfides (olive green).

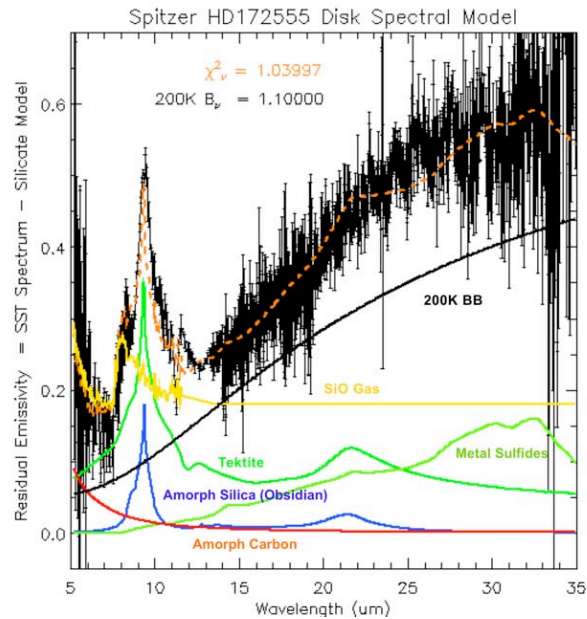

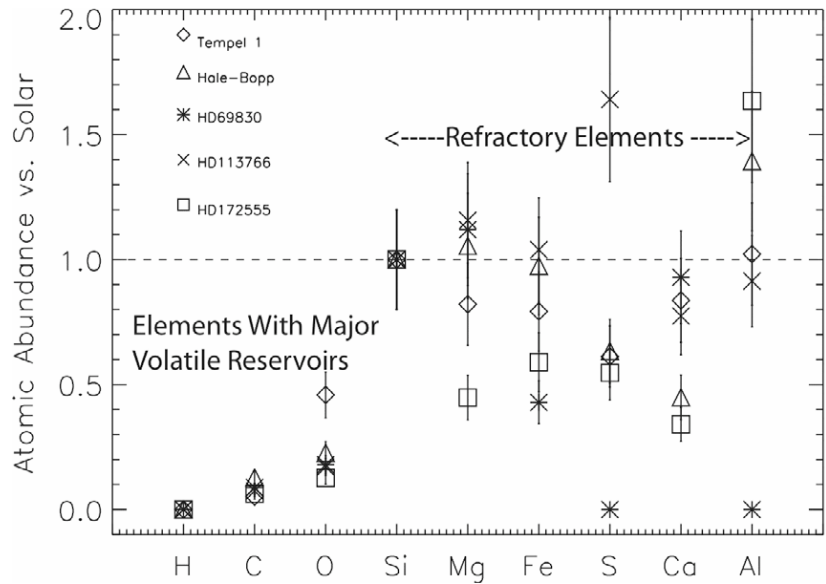

**Figure 5** - Derived elemental abundances for the HD172555 dust excess, relative to solar abundances, and as compared to other Spitzer dust spectra (Figure 2). The Si relative abundance has been set = 1.0. The major refractory species, with the exception of Al. are consistently depleted versus solar. If the contribution of atoms from the detected SiO gas is included, all atomic species decrease in abundance versus solar by approximately a factor of 2, except for O, which increases to approximately 0.6 (Table 3).

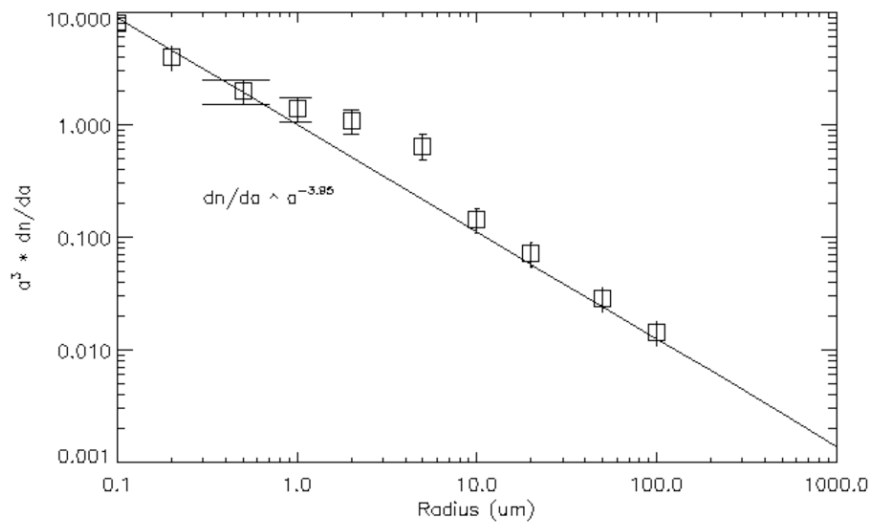

**Figure 6** - Derived particle size distribution for the HD172555 dust excess producing the strong silica and silicate emission features. Error bars are estimated conservatively at 25% of the relative abundance at a given size. The 200K large particle population can be represented on this figure as any combination of particles with sizes ≥100 μm (0.1 mm) and total surface area ~80% of the area in the fine dust PSD.

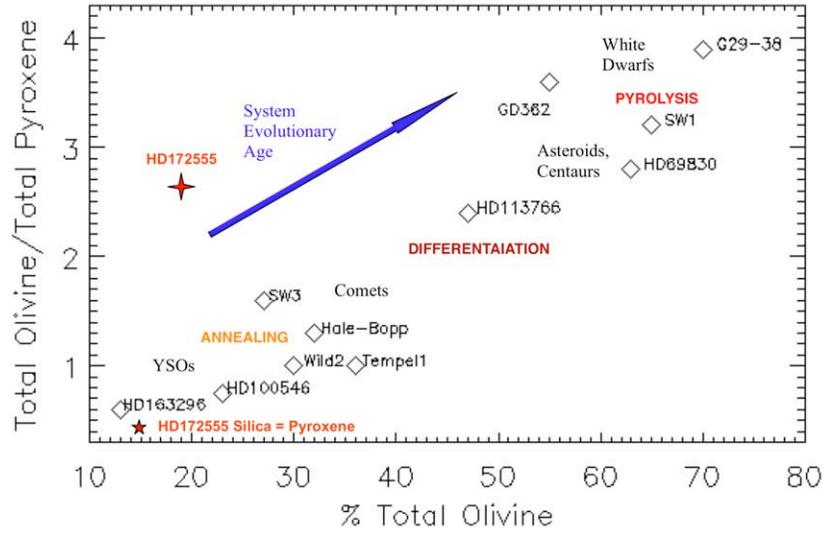

**Figure 7** — **Silicate mineralogy for HD172555** versus that found in 4 comets systems (SW3, Sitko, *private communication* 2007; Hale-Bopp, Lisse *et al.* 2007a; Wild 2, Zolensky *et al.* 2007; and Tempel 1, Lisse, C. M. *et al.* 2006), the primitive YSO disk systems HD100546 (Lisse, C. M. *et al.* 2007a) and HD163296; the young F5 terrestrial planet forming system HD113766 (Lisse *et al.* 2008); the mature asteroidal debris belt system HD69830 (dominated by P/D outer asteroid dust; Lisse, C. M. *et al.* 2007b); the Centaur SW-1 (Stansberry *et al.* 2004), and the ancient debris disk of white dwarfs G29-38 (Reach *et al.* 2008) and GD362 (Jura 2006). The general trend observed is that the relative pyroxene content is high for the most primitive material (i.e., YSOs), and low for the most processed (i.e., white dwarfs). HD172555 is far off this trend line (blue star). If we assume all the detected silica species were originally pyroxene, then the HD172555 point lies on the trend line in the YSO region of the plot (red star).

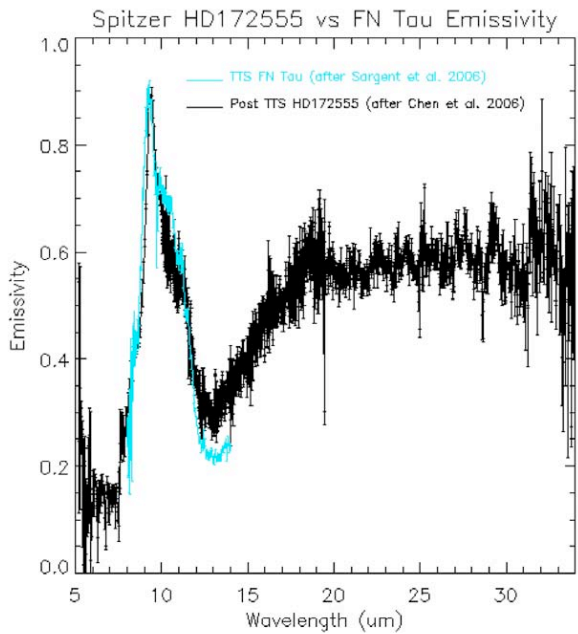

**Figure 8** - Comparison of the emissivity spectra of the A5 star HD172555 (Chen *et al.* 2006) with the published 8-13 μm emissivity spectrum of the $< 10^7$ yr old, 0.1 $M_\odot$ T Tauri system FN Tau (Sargent *et al.* 2006), recently shown to have a near face-on, optically thick disk by Kudo *et al.* (2008). The match of the silica and silicate emission features is surprisingly good, suggesting that the hypervelocity collision mechanism found to be operative in the HD172555 system may also be functional in T Tauri systems.